# On the frequency and temperature dependence of the conductivity

Çağlar Tuncay
Department of Physics, Middle East Technical University
06531 Ankara, Turkey
caglart@metu.edu.tr

**Abstract**: We obtain the frequency and temperature dependence of the conductivity for (disordered) solids, where the temperature dependence is defined in terms of the related thermodynamic state function in exponential (or power law, etc.) forms. The model is applied to n-type Si with various donor and acceptor impurities and several concentrations at $T \sim 0\ ^0K$ and sodium borate glasses ($xNa_2O \cdot (1-x)B_2O_3$) and mixed alkalis ($xLiF–(0.8-x)KF–0.2Al(PO_3)_3$) with various compositions (x) in all at $T \sim 500\ ^0K$. The results are found in good agreement with experiments.

**1 Introduction:** It is known that, the frequency dependent conductivity ($\sigma(w)$) displays power law ($\sigma(w) \propto w^p$) for many disordered solids [1, 2], if w is big yet less than phonon frequency and p is about one. For several materials and several models; see, [3-5] for "ion conduction", and [6, 7] for "electron conduction", and [8] for a highly viscous ionic melt just above the glass transition. Many more examples about the subject, which is known as Taylor-Isard scaling [9], may be found in the given reference list. The following issues may be distinguished within the known theories (the symmetric hopping model, the variable range hopping model, the effective-medium approximation, the percolation path approximation, the diffusion cluster approximation and others [2, 3, 10]: i) The local free charge conductivity is Arrhenius temperature dependent (i.e., $g = g_0 \exp(-E_{act}/k_B T)$ where g and $E_{act}$ denote the free charge conductivity and the activation energy, respectively. [11] ii) Conduction may be classical or quantum mechanical. [12] iii) In terms of discrete Maxwell equations, the inhomogeneous solid is approximated by a RC network, where the resistor (R) currents are the free charge currents and the capacitor (C) currents are Maxwell displacement currents. So, each resistor is proportional to the inverse free charge conductivity and all capacitors are equal. [13] iv) For high frequencies, the (common) admittance of the capacitors in the network dominates and the electric field becomes spatially homogeneous, where the average resistor current is the average free charge conductivity. It is widely accepted that, percolation [14] is underlying ac universality in the extreme disorder limit. See Dyre, [13]; and also references in [14] for the related theoretical review articles. Reviewing the literature and discussing the known theoretical approaches are kept beyond the scope of this contribution. We present a new model. Our essential aim is to show that, many aspects of the ac universality for disordered solids maybe emerging out of simple relations between the conductivity and the frequency and temperature, which may be described in terms of familiar concepts of the electromagnetic theory (Maxwell equations).

We followed three approaches: We solved random LRC network in terms of the current paths (please see Appendix); we solved the loop equations (Maxwell equations) and finally we applied Maxwell theorem (equipartition of energy theorem). We obtained the same results for conductivity out of the mentioned approaches. We follow the loop theorem i.e., Maxwell equations here. The following section is the model. The next section is for the application of the model to: a) The n-type silicon with various acceptor and donor impurities with several concentrations at $T \sim 0\ ^0K$ [15]. b) The glasses; sodium borates $xNa_2O \cdot (1-x)B_2O_3$ [16] and mixed alkalis $xLiF–(0.8-x)KF–0.2Al(PO_3)_3$ [3] at $T \sim 500\ ^0K$ with various compositional concentrations (x), in all. The last section is devoted for discussion and conclusion. Appendix is for some details of the model. Please note that, we follow SI units.

**2   Model:**   The model is composed of two original models; one for the frequency dependence of the conductivity at a given temperature (Sec. 2.1) and one for the temperature dependence of the conductivity for a given frequency (Sec. 2.2).

*2.1 Frequency dependence:* Electrical conductivity is known as a matrix ($\sigma_{xy}$) connecting the components of the vectors for the applied electric field ($E_y$) and the resulting current density ($J_x$),

$$J_x = \sigma_{xy} E_y \quad . \tag{1}$$

The conductivity may be considered as a scalar quantity, if the disorder in the sample is homogeneous (and isotropic with respect to the direction of the electric field, up to randomness), etc.;

$$J = \sigma E \quad . \tag{2}$$

where, J and E are approximate or the average values (over many similar cross sectional areas for J and paths for E, respectively): I=∫**J**.d**a** and emf=∫**E**.d**l**, where emf is the electro motive force.

We think that there are three contributions (in the main terms) to the current density:

1) Capacitive contribution: The ac conductors are inserted as dielectric slab between the electrodes (probes) of the ac power supply; so that the solid becomes a capacitor with $C=\varepsilon\Delta/\delta$ (if the side leakages of the electric field, i.e. the fringe effects are ignored), where $\varepsilon$ represents the material (chemical, structural, etc.) properties of the solid and $\Delta/\delta$ is the geometrical (shape) factor; and $\varepsilon$ is defined in terms of the permittivity constant for free space ($\varepsilon_0=8.85\times10^{-12}$ Farad/m) and the relative permittivity (dielectric constant of the solid) as $\varepsilon=\varepsilon_r\varepsilon_0$, which is known to be a complex number. For a parallel plate capacitor, $\Delta$ is the cross sectional area and $\delta$ is the distance between the plates (thickness) of the capacitor. We have displacement currents (Maxwell equations) between the plates of the capacitor and current is spatially constant; so that one may define a current density as a function of time as $J_C(t)=I_C(t)/\Delta$, where $I_C(t)$ is related to the charge ($Q_C(t)$) of the capacitor; $Q_C(t)=-\int I_C(t)dt$. More over, if a complex exponential function is defined for the ac current ($I_C(t)=I_0\exp(iwt)$) with $I_C(t=0)=0$, where $I_0$ is the amplitude for the displacement current and $i=(-1)^{1/2}$), then $Q_C(t)=(-1/iw)I_C(t)$. As a result;

$$J_C = \varepsilon w E \quad , \tag{4}$$

where, the complex ($\varepsilon$) is known to be w-dependent which may also depend on T.

2) Resistive (Ohmic) contribution: It is known that the resistive (i.e., conduction) current density ($J_R$) is defined as (for a given field, microscopic Ohm law);

$$J_R = (1/\rho) E \quad , \tag{5}$$

where ($\rho$) is the resistivity of the solid, which is known to be a real number for a given T. There is no evidence about the frequency dependence of ($\rho$). It may be that ($\rho$) depends on w (as ($\varepsilon$) does), yet the mentioned relation may not be recognizable within the experimental frequency range for the ac. It is obvious that, Eq. (2) is approximately (or as an average) valid in the present case.

3) Inductive contribution: We think that, we have inductance (self and mutual) in

disordered solids. It is known that [11, 17-19] the charge carrier displacement has a negative phase with respect to the electric field and the introduced lag is at most quarter of a period; so that (the current reaches its maximum before the field, and) we have a capacitive response rather than an inductive one. For that reason, inhomogeneous solids are considered as RC networks by many authors (see [17], for example). But, the introduced lag is not exactly ¼ of a period; and we do not ignore L (some other reasons are discussed in Sec. 3 and 4 and Appendix). On the other hand, the charge neutrality and the principle of energy minimization must be satisfied locally or extensively in the solid, where it is well known that the host charges are distributed in alternating fashion over the host sites (lattice). As a result, one may expect that the extremes of the potential for disorder are distributed spatially homogenous and they alternate in any direction: At the preparation (formation) temperature of the solid, the highest (lowest) potential centers (regions, with respect to the distribution of the potential of the host crystal field) must be followed by low (high) potential ones, along any direction. Conclusively; the polarization (at low T) or the conduction (at high T) currents may follow spiral paths as well as straight ones. In any case, we may take inductances (solenoids) in ac solids, where the induced electromotive force is known to be the multiplication of the inductance and the time derivative of the current ($dI(t)/dt=iwI(t)$). The physical definition of the inductance for a solenoid ($L_{sol}$) of length ($\delta$) is known to be $L_{sol}=\mu n^2 a\delta$ (for a section of some length near its center or if the side leakages of the magnetic field are ignored), where a is the cross sectional area of the solenoid (which is smaller than or at most equal to $\Delta$), n is the number of the turns per length (wounding for magnetic flux linkages which may be proportional to the number of disorder centers per length along the ac electric field) and $\mu$ is the permeability of the solid, which is equal to the multiplication of the relative permeability of the solid ($\mu_r$) and the permeability for free space (vacuum, $\mu_0=1.26 \times 10^{-6}$ Henry/m). It is known that, ($\mu_0$) and ($\epsilon_0$) are related in terms of the speed of light in vacuum ($c_0=(\mu_0\epsilon_0)^{-½}$). Similarly, the speed of light in a medium (c) is related to $\mu$ and $\epsilon$; and, the ratio ($c/c_0$) defines the optical index of the solid with respect to vacuum. In short, we have inductance in the ac conductor, where the out coming electromotive force is $emf_L=-LdI_L/dt$ and $emf_L$ is known as the line integral of the electric field along the induced current. It is obvious that, Eq. (2) is approximately (or as an average) valid in the present case. More over, we expect that the inductance(s) in disordered solids may be similar to solenoids and we have many inductances in a solid (close packed inductances) and mutual inductance between them must also be taken into account for the inductive contribution to the currents in a solid (please see Sects. 2.3, 4 and Appendix.)

Combination of the mentioned equalities for the inductance (solenoid) yields

$$J_L=I_L/a=(g/i\mu w)E \quad , \tag{6}$$

where, g is the length (g', say) of the path for the current in an individual inductance (which is greater than or equal to the length of the sample along E) divided by La/$\mu$; g=g'$\mu$/aL. (We do not know the type (shape) of the inductance; yet, for any shape $L=\mu n^2 g''$, where n is the number of wounding per length and g'' is the related geometrical (structure) factor. (For example, g''=a$\delta$ for a solenoid with cross sectional area a and length $\delta$.) Thus, g=(g'/a)/($n^2 g''$) for any inductance, where g is real.

Now, we may sum the current densities for the total current (Eqs. (4) to (6)) for the applied ac electric field at a time t, where the organization of the out coming equation reads

$$\sigma(w)= g(i\mu w)^{-1} + \rho^{-1} + i\epsilon w \quad . \tag{7}$$

In Eq. (7) $\mu$ and $\varepsilon$ are complex parameters and they depend on frequency and temperature; yet, they may be taken approximately as constant in some w and T regions. The real part ($\sigma_R(w)$) of the conductivity (Eq. (7)) for $\varepsilon=\varepsilon_1 - i\varepsilon_2$ and $\mu=\mu_1 - i\mu_2$ is

$$\sigma_R(w)= g(\mu_3 w)^{-1} + \rho^{-1} + \varepsilon_2 w \quad , \tag{8}$$

where, $\mu_3=\mu_{mod}/\mu_2$ with $\mu_{mod}=(\mu_1^2+\mu_2^2)$; and g pronounces size effect in the electrical conductivity of the solid, which is related to the inductances.

We drop the sub indices 3 and 2 from the parameters $\mu$ and $\varepsilon$, respectively ($\mu_0 \rightarrow \mu$, $\varepsilon_2 \rightarrow \varepsilon$); and change the notation in Eq. (8) as; $\sigma_R \rightarrow \sigma$ (i.e. we will use from now on $\sigma$ for the real part and $\sigma_{im}$ for the imaginary part of the conductivity) as followed widely in literature:

$$\sigma(w)= g(\mu w)^{-1} + \rho^{-1} + \varepsilon w \quad . \tag{9}$$

This finishes the definition of the model for the frequency dependence of the electrical conductivity in disordered solids. Some of the related subjects may be found in Secs.: 2.3, 3, 4 and Appendix. The next section is the temperature dependence of the conductivity versus frequency.

*2.2 Temperature dependence:*
It is widely accepted that the conductivity is mainly due to polarization currents (besides the displacement currents, which are always present) at $T \sim 0\ ^0K$ and the conduction currents become important as T increases. Secondly, the physical dimensions of the experiment samples vary with T and thermal expansion may be important for the following two cases: If the experimental temperature range is big and if the inductance of the sample is important for the conduction, since the geometrical factor for the sample enters the conductivity as a coefficient for the inductive contribution (Eq. (9)), where the thermal expansion may also be important for $\mu$, $\rho$ and $\varepsilon$. Yet, we think that the inductive term in the conductivity is negligible for big T, where the size effect (which is pronounced in terms of the g-factor) is also negligible (since L is big). But for $T \sim 0\ ^0K$, the inductive currents and hence the size effects may become very important, as we consider in several sections and Appendix.

Thermodynamic state function is known to yield the variation of the volume (Vol=$\Delta\delta$) of a material (in a certain phase) with temperature (T) under a given pressure: dVol/dT=f(Vol,T,$T_0$), which may be written as a sum of the terms $\xi Vol^\eta T^\zeta$ with various $\xi$, $\eta$ and $\zeta$ about some given temperature ($T_0$). In several cases f(Vol,T,$T_0$) may be approximated by a single term $\xi Vol^\eta T^\zeta$ (instead of a sum), where the parameters for $\xi$, $\eta$ and $\zeta$ may be taken as constant in some temperature regimes (for a given phase) about some $T_0$ (if the thermal expansion is not abrupt, etc.). It is obvious that $d\log_e Vol/dT^{\zeta+1}=\xi(\zeta+1)^{-1}$ for $\eta=1$ and we have stretched exponential for the variation of Vol in T, which is appropriate for liquids; whereas, if $\eta \neq 1$ then we have power functions (or expansion in a power series) for Vol in T, which is useful for gasses (in adiabatic processes, for example), yet it may be suitable for some ac solids in several frequency and temperature regions as well.

We may assume similar (stretched exponential or power law) expansions in T for $\rho$ and $\varepsilon$ and it is known that the exponential expansion of the volume and the resistivity in T well suits for many solids; so that we follow the same approach for $\varepsilon(T)$. In summary, we assume

$$Vol(T)=Vol(T=T_0)\exp(\alpha(1+\gamma)^{-1}(T^{\gamma+1} - T_0^{\gamma+1})) \tag{10}$$

for the volume and similarly for $\rho$ and $\varepsilon$ (and $\mu$), where one or two power terms in T may be added into the exponent of Eq. (10) for some ac conductors in some temperature regimes and utilize Eq. (10) for $\Delta(=Vol^{2/3})$ and $\delta(=Vol^{1/3})$ in the disorder and size factor (g in Eq. (9)), etc. For example for $\rho$, we take the well known Arrhenius behavior and merge it with the behavior for the thermal expansion; please see Sec. 3 for the applications. Hence, we have one or two more terms, besides Arrhenius (activation energy) temperature behavior, in the exponent for $\rho$. This finishes the definition of the model for the temperature dependence of the electrical conductivity in disordered solids, for a given frequency. Some of the related subjects may be found in Secs. 2.3, 3, 4 and Appendix.

*2.3 Supplementary:*

1) The (real part of the) conductivity (Eq. 9) is *dynamic* since it is connected to $\epsilon$ which is frequency dependent and the imaginary part of the complex conductivity ($\sigma_{im}$) and hence the total dielectric function may be obtained in terms of Kramers-Kronig transformation. (The imaginary part of $\epsilon$, i.e., the dielectric loss, is the same as the imaginary part of the susceptibility where the susceptibility is equal to the magnitude of the polarization field per magnitude of the electric field per $\epsilon_0$.) Similarly, $\mu$ (and $\rho$, as well; see case 2) in Sect. 2.1) may be frequency dependent.

Some intuitive, phenomenological or empirical expressions similar to $\sigma(w)$ in Eq. (9) may be found in the literature; for example, Pollak and Geballe utilize an expression (Eq.1 in [15]) (for the conductivity in n-type silicon with various impurities) where the frequency dependence displays a power law, which is less than unity and varies from sample to sample; and, the coefficient for the mentioned power function changes with temperature (Table II in [15]). Elliot defines a similar power law for the frequency dependent conductivity (Eq. 1 in [17]). Kulkarni *et. al.* approximate their empirical data for mixed alkali glasses with the sum of a power law and the "dc conductivity" [3]. Some analytical expressions similar to Eq. (10) here may be found in [2] for the temperature behavior of R(T) and C(T); Eq. (10) there.

2) We have three regions in $\sigma(w)$ for a given T: $\sigma(w)$ increases with increasing w (for big w) in the right region (branch, arm) where the capacitive term (Eq. (9)) is important (C-branch); $\sigma(w)$ is almost horizontal for intermediate w in the middle region (valley) where the resistive term (Eq. (9)) is important (R-valley) and $\sigma(w)$ increases with decreasing w (for small w) on the left region (branch, arm) where the inductive term (Eq. (9)) is important (L-branch). Conductivity for small w (due to the inductive term in Eq. (9)) may become important for big g; i.e., for long and thin samples as rods, tubes, nano tubes, etc. The conductivity follows power law unity asymptotically (w→∞). The *minimum* of the conductivity at a T ($\sigma_{min}(w)$) defines a special frequency $w_0=(g/\mu\epsilon)^{1/2}$ which is the natural frequency for oscillation in the L(R)C circuit (resonance frequency) and obviously related to the optical index of the sample ($w_0$ is the frequency where the derivative of the conductivity (Eq. (9)) with respect to w is equal to zero): $\sigma_{min}=\sigma(w_0)=2\epsilon w_0 + \rho^{-1} = 2(g\epsilon/\mu)^{1/2} + \rho^{-1}$ and $\sigma_{min} \neq \rho^{-1}$. Hence, $\sigma(w=0)$ is undefined and $\sigma_{min} \to \rho^{-1}$ as $w_0=(LC)^{-1/2} \to 0$, i.e., as L→∞ or C→0. (There may be no dc at all; and all the "dc generators" may be supplying currents with several oscillations with various frequencies about $w_0$, which is known as noise.) For the *maximum* of the $\sigma(w)$, it is claimed that (Dyre, [in 13 and 18]; also [19]), $\sigma(w) \propto w^p$ behavior survives up to phonon frequencies (where the conductivity around $w=10^{12}$ Hz is of order 1 $(\Omega cm)^{-1}$) and there various resonance phenomena take place so that $\sigma(w)$ decays to zero sharply. We estimate $\sigma_{max}=\sigma(w=10^{12}$ Hz$)=\rho^{-1}+\epsilon 10^{12}=\rho^{-1} +\epsilon_r\epsilon_0 10^{12} \cong \epsilon_r(8.85 \times 10^{-12})10^{12} = 8.85\epsilon_r$ $(\Omega m)^{-1}$, where $\epsilon_r$ is same as in Sec. 2.1 case 1), which is less than ten for many solids (for example $\epsilon_r \sim 5$ for Silicon and pyrex glass). And we have $\sigma_{max} \cong 0.45(\Omega cm)^{-1}$ for many inhomogeneous solids. Furthermore, at $w=10^{12}$ Hz (i.e. at about the frequency for the maximum of the conductivity) $\hbar w \cong k_B T$, where $\hbar(=h/2\pi \cong 10^{-34}$ J-sec) is Planck constant and $k_B$ (=1.38x10$^{-23}$ J/$^0$K) is Boltzman constant; $\hbar w \cong 10^{12} 10^{-34}=10^{-22} \cong 1.38 \times 10^{-23}$T, even at small temperatures.

Hence, $w \cong k_B T/\hbar$ (phonon frequency) is a threshold for the quantum effects (photons) to become important in the solid; in other words, the phonon frequencies may be considered as the quantum frequencies ($w_{quantum}$) and for $w < w_{quantum}$ the underlying physics for the conductivity is classical (Maxwell equations).

3) The *relaxation* time constants for RC and LRC ac circuit elements are known as $\tau_C = RC (= \rho\varepsilon$, for parallel plate capacitors and Ohmic resistors) and $\tau_L = L/R (\propto \mu/\rho)$, respectively. Hence, Cw and 1/Lw have the same dimension; per resistance (where the conductivity has the dimension per resistance per length). Secondly, $\tau_L \tau_C = LC = w_0^{-2}$, where $w_0$ is same as in Sec. 2.3; case 2). Thirdly, $\tau_C \ll \tau_L$ and inductive relaxation currents decay more quickly than the capacitive ones in thin and flat solids. Finally, Eq. (9) can be organized for $(\sigma(w) - 1)\rho = g((\mu/\rho)w)^{-1} + \varepsilon\rho w$ which may be written in terms of the relaxation time constants:

$$(\sigma(w) - 1)\rho = (\tau_L w)^{-1} + \tau_C w = P/(2\pi\tau_L) + 2\pi\tau_C/P \quad , \quad (11)$$

where $P = 2\pi/w$, i.e. the period of the ac and $(\sigma(w) - 1)\rho$ is dimensionless. More over, since $\mu_0 \propto 1/\varepsilon_0$ and temperature dependence of one of g, $\mu_0$, $\varepsilon_0$ or $\rho$ (Eq. (10)) may be obtained from this of another one (in terms of scalar multiplications, i.e., scaling the temperatures, etc.,) at least in some certain temperature regimes and approximately, then the conductivity in Eqs. (9) or (11) or the experimental data may be obtained as the same or collapsed onto a single curve, i.e. the so called "master equation" or "master curve", respectively; for various solids in various frequency and the temperature regimes which is known as the "universality" in the ac.

4) The conductivity (Eq. (9), where $\mu$, $\varepsilon$ and $\rho^{-1}$ have the dimension Henry/m, Farad/m and Ohm/m, respectively and all real) may be written in terms of the parameters for the ac circuit elements (L, R and C) defined for the solid and the corresponding geometrical coefficients ($g_L$, $g_R$ and $g_C$)

$$\sigma(w) = g_L (Lw)^{-1} + g_R R^{-1} + g_C Cw \quad , \quad (12)$$

where $g_L$ is equal to the length of the path for the current passing through the inductance per cross sectional area for the same current and $g_R = g_C = \Delta/\delta$. Eq. (12) pronounces the size effect for the frequency dependent conductivity.

5) It is natural that we have three terms in the conductivity (Eq. 9) each of which is for one of the ac circuit elements; L, R or C and this phenomenon is not special to ac conductors: Any electrical element (material) involves some inductance, some resistance and some capacitance at the same time; as it is well known. In other words, any conductor is a resistor, a capacitor and an inductor at the same time, where the inductive character emerges in terms of the interacting current paths in the material, etc. Yet, the underlying current mechanisms may be different at different temperatures and frequencies for the ac. For example, the charge carriers may reach from one electrode to the other within a period of time equals to $2\pi/w$ if w is small. But, for $w = 10^6$ Hz or big, the related drift velocity ($v_d \sim w\delta/2\pi$) may be estimated as about 1000m/s or big for a thickness $\delta \sim 6$ mm, which is not what we expect. Thus, the underlying currents must be of displacement or polarization type (since charges oscillate locally), for high frequencies ($10^6$ Hz$<$w). Similarly, for low temperatures (about the absolute zero), we do not have many free charge carries in the solids; hence, we may expect the conductivity as emerging in terms of the displacement or polarization currents, where L may be finite since we have a limited number of channels for the conduction currents at a time if the disorder is not extreme. For $T \propto E_A/k_B T$, i.e., at about the ionization temperatures, conduction currents also contribute the conductivity.

Another source for the induced magnetic fields in the ac solids is known as the time variation of the electric field of the applied ac (due to Faraday-Lenz law, Maxwell equations). As a result we have electric and magnetic fields within the ac solids (D and H respectively) or slabs between the plates of a capacitor as dielectrics at the same time, where it is known that H~D/c and H is ignored. But this ignorance is in contradictive: The energy stored per volume (u=U/Vol) of the material is the sum of the terms $u_D=½εD^2$ and $u_H=½H^2/μ$, where none of the terms may be omitted due to a well known Maxwell theorem which is called the equipartition of energy. The mentioned sum may be conserved if there is no resistance (dissipation) for the currents, or vice versa. The total energy due to the electric and magnetic fields ($U_D$ and $U_H$, respectively) at any (point and) time are given as $U_D(t)=½Q^2/C$ and $U_H(t)=½LI^2$. Hence, we have inductance, due to theorem of the equipartition of energy, since we have capacitance. The resistive (Ohmic) term is known as dissipative. Thus, one may utilize Maxwell energy equipartition theorem together with the energy conservation theorem and obtain conductivity (Eq. (9)).

The next section is the application of the model to various disordered solids in various temperature regimes.

## 3    Application and results:

In some experimental reports the physical dimensions of the sample and some information about ρ are given, yet there is little information about ε (so that C could be estimated in terms of the physical dimensions) which is indeed dσ(w)/dw for big w and there is no information about μ (or L) at all. Also, there is no empirical data about the thermodynamic state functions and the related exponential coefficients for thermal expansion (contraction). We predict the model parameters by utilizing the given empirical data for σ(w,T) at certain w or T values; and then, we utilize these parameters for different w or T (in terms of extrapolation). It is obvious that, we do not follow fitting at all (yet it could be done to define all the related parameters in w or T, etc., for better extrapolations): We solve the equations at some points for σ(w) and mimic the empirical data and try to make predictions. Below are the results with the parameters given in Tables I, II and III for doped Si [15], sodium borate glasses [16] and mixed alkali glasses [3], respectively.

Various discussions about the model are given in further sections and Appendix; and, it may be worth to brief the underlying currents in several temperature regimes here: We have mainly displacement and polarization currents at small temperatures; $T<E_A/k_B$ where $E_A$ is Arrhenius (activation) energy. Conduction currents gain importance at intermediate temperatures, where $T\sim E_A/k_B$. We think that the polarization and the conduction currents follow spiral paths along the disorder centers between the electrodes. Please note that, we always have displacement (Maxwell) currents and that is why the capacitive term is important for the conductivity (Eq. (9)) for $w_0<w$ (the experimental frequencies).

*3.1. n-type silicon with various acceptor and donor impurities with several concentrations*:

Several experimental information may be found in [15] about the experiment samples (Si), where many majority (n-type) and minority (p-type) impurities with several concentrations have been incorporated into Si by various fabrication methods, which are all given in Table I, *ibid*. The physical dimensions of the sample are given on page 2: "The samples were disk-shaped, about 1 cm$^2$ in cross section and 1/3 mm thick, so that conductivities of the order of $10^{-16}$ w could be meaningfully measured." Hence, the parameter for the g-factor in Eq. (9) could be easily estimated (for a given inductance type) by assuming that the impurities (Table I, *ibid*.) are uniformly (random) distributed, but we do not know the real and complex parts of μ for the experiment samples. On the other hand not g but g/μ is important in σ(w) (Eq. (9)). The inductance for an individual solenoid (with cross sectional area a∝Δ/N') in Si may be estimated in terms of the definition: $L_{sol} \cong \mu n^2 Vol$, where Vol is the volume of the solenoid ($=\Delta\delta$) and n is the number of turns in spirals per length ($n=N/\delta$ with $NN'=M$, where M is the total number of the impurities). We think that the recent estimation may be valid and important for the conduction in long and thin solids (rods, wires, nano tubes, etc.). It may be shown that, $L_{sol} \cong \mu(N^3/M)\Delta/\delta$ for a single solenoid. Whereas, in the solid we have many of them (close packed inductances) and mutual inductance between them must also be taken into account for the inductive contribution to the conductivity.

At high temperatures or at the extreme disorder limit, where the number of flux linkages is big, the currents in loops or in solenoids overlap (due to mutual attraction of the parallel currents) and unify to bigger loops or wider solenoids. If we consider the whole solid as a single solenoid (where we should have surface currents on the (lateral) surface(s) parallel to the direction of the ac), then we have $L_{sol}=\mu(N^2/\delta^2)\Delta\delta$, where $N \cong M^{1/3}$ and M is equal to Vol times the concentration ($\chi$) of the disorder (concentrations are given on the last two columns of Table I *ibid.* for various n-type Si). Hence, $L_{sol} \cong \mu\chi^{2/3}\Delta^{5/3}\delta^{-1/3} \sim 100\mu_r$ Henry, where the donor (Phosphorus) concentration ($\chi=1.5\times10^{16} cm^{-3}$) of the first sample (no=8 and crystal no A78-25A) *ibid.* is used and we do not know how correct our estimation is, since we do not know whether solenoid model is valid here (there is no available information about $\mu$, L, etc.). If we knew the relative permeability ($\mu_r$) of the given n-type Si, then we could estimate L for the sample more precisely. We will try to predict $g/\mu$ empirically for several samples, where we will utilize the related experimental $\sigma(w,T)$ plots given Pollak and Geballe [15].

For the definition of the parameter for $\varepsilon$ (or C) we utilize Fig. 5 *ibid.* from which we reproduced Figure 1 here. We observe that $log\sigma(w)$ is about $8\times10^{-9}$ $\Omega^{-1}cm^{-1}$ for $w=100KC(=10^5)$, where w is big enough and the inductive contribution to the conductivity may be ignored for $T \sim 0$ $^0K$ where Arrhenius behavior (and thermal expansion, which may be minor $T \sim 0$ $^0K$) gains importance as T increases. We assume that $1/\varepsilon \ll \rho$ and we estimate $\varepsilon \cong 8\times10^{-12}$ Farad/m (with $\varepsilon_r=\varepsilon/\varepsilon_0 \cong 0.9$) for $T \sim 2$ $^0K$, where C may be estimated as $24\times10^{-12}$ Farad since $C=\varepsilon\Delta/\delta$ (for a parallel plate capacitor), where $\Delta=1cm^2$, and $\delta=1/3$ mm, *ibid*. In Figure 12 there, (which is about the measured capacitance of the sample plus holder including frequency-independent parts, as noted in the caption of the recently mentioned figure) the capacitance is given as about $23.5\times10^{-12}$ F at $T \sim 2$ $^0K$ for 100 Hz<w. Hence, we think that our estimation for $\varepsilon$ (C) may be considered as meaningful.

More over, we observe in the same figure(s) that the conductivity $\sigma(w)$ behaves differently at different temperatures: At $T \sim 2$ $^0K$ or smaller, $\sigma(w)$ and hence $\varepsilon$ is almost independent of T (or 1/T). The parameter for $\varepsilon$ (C) varies smoothly in T, since each plot stays (almost) horizontal till $10/T \sim 4$. For $10/T \rightarrow 0$, the derivative of $log\sigma(w)$ with respect to $10/T$ is constant which is independent of w, and it is nearly -3 as the inclined arrow (dashed) indicates. Therefore, we have a different temperature regime for $\sigma(w)$ now, which is not solely capacitive. More over, the conductivity increases abruptly (by a factor about hundred or bigger) as T increases by few Kelvin there. The mentioned temperature region is known as the ionization range as noted by Pollak and Geballe on page 3 in [15]. ("The measurements were made over a temperature range from 1.2 $^0K$ up to the ionization range where conduction band electrons obscured the impurity conduction, usually about 25 $^0K$.") We share the idea of Pollak and Geballe that, the polarization may not follow the ac for high w. Secondly; as T increases, more ionization takes place and at $T \sim E_A/k_B \sim 25$ $^0K$ or bigger, the conduction currents gain importance. The conductivity for the conduction currents has the slope -3 in logarithmic for ($\sigma$) and linear (in 10/T) scales. Hence, $dlog_{10}\sigma(T)/d(10/T) \sim -3$ and the exponential coefficient for resistivity in Eq. (10), i.e., $\alpha_\rho(1+\gamma_\rho)^{-1}$ may be taken as about -30 for $\gamma_\rho=-2$; hence $\alpha_\rho=-30$. So we take $T_0=1.2$ $^0K$ and the temperature dependence of the resistivity becomes

$$\rho(T) = \rho(T_0=1.2)\exp(-30(T^{-1} - 1.2^{-1})) \qquad (13)$$

which is valid only for the given doped Silicon (the caption of Fig. 1 here) and for $T<25\ ^0K$. For other impurities in Si, or at a different temperature regime, only a small variation in the exponential coefficient may suffice for a small difference in Arrhenius activation energy ($E_A$) in $\rho(T)$ and hence in $\varepsilon(T)$ and $\sigma(T)$, etc. For other cases, the number of the temperature terms with various powers and the exponential coefficient may be different; where, the known theoretical approaches (such as the symmetric hopping model, the effective-medium approximation, etc. as mentioned in Sec. 1 here) may be useful to predict the temperature dependence of the resistivity and other parameters rather than the frequency dependent conductivity.

We may refine Eq. (13) by adding some terms in T (and $T_0$) with different power which takes care of the thermal expansion at about $T\sim 0\ ^0K$:

$$\rho(T) = \rho(T_0=1.2)\exp(-34((T^{-1} - 1.2^{-1}) + (T^{-\frac{1}{2}} - 1.2^{-\frac{1}{2}}))) \quad, \qquad (14)$$

where we predict that the resulting expansion in a physical dimension (radius or thickness, say) is much less than 1%, which is hardly noticeable since the physical dimensions of the experiment sample are small.

Please note that, Fig. 5 *ibid.* or Fig. 1 here shows variation in the envelope for the conductivity (left up corner) for big T and at about $\sigma=5\times10^{-5}\ \Omega^{-1}cm^{-1}$ (for about $w=10^9$ Hz). We make the following claim about the mentioned change in the conductivity: For $25\ ^0K<T$, many electrons are ionized and conduction currents are important in the present temperature regime (for $25\ ^0K<T$), where the individual inductive currents (in terms of polarization) decayed and hence, the whole solid became a single inductance (solenoid), where the individual inductive currents in loops unify. We may represent the mentioned cross over in the conductivity by adding a small term in $(T-T_0)$ into (14), which becomes important for big T ($25\ ^0K<T$):

$$\rho(T) = \rho(T_0=1.2)\exp(-34((T^{-1}-1.2^{-1}) + (T^{-\frac{1}{2}}-1.2^{-\frac{1}{2}}) - 0.004(T-1.2))) \quad. (15)$$

The other related parameters are given in Table I here (the first row) and the resulting plot is displayed in Figure 2.

Figure 3 is for the conductivity isotherms versus w, where the range for the frequencies is (intentionally) much wider than the one used in experiments. The temperature increases in Fig. 3 from $1\ ^0K$ (thick solid line) to $20\ ^0K$ (thick dashed dotted line) in equal steps ($1\ ^0K$) for the plots from bottom to top. The inset is for the reciprocal of $\rho(T)$ (Eq. (15)) which shows that $\sigma(T)$ and $\rho^{-1}(T)$ are dissimilar. Further more, we do not call $\rho^{-1}$ the dc (synonymously the frequency independent) conductivity or the conductivity $\sigma(w\to 0)$: Because, we think that, $\rho^{-1}$ is equal to neither $\sigma_{min}$ nor $\sigma(w\to 0)$, where $\sigma(w\to 0)$ is not defined at all, as given in Fig. 3. Secondly $\rho^{-1} \neq \sigma(w=w_0)$ for $T<E_A/k_B$, but $\rho^{-1}=\sigma(w=w_0)$ for $E_A/k_B<T$ as we had already discussed theoretically in Sect. 2.3 case 3). We consider the subject on the experimental basis within the following paragraphs.

*Arrhenius behavior, the minimum of the conductivity and the inductance:* We pay a special attention to the plots for small conductions in Fig. 1, where we have decrease in the conductivity in almost equal steps about unity as horizontal and two sided arrows below the one for w=100 indicate. The smallest conductivity given in [15] for the dc seemingly corresponds to w=0.01. There is a note on page 4 *ibid.*, which states that, "The reduced slope at the low-temperature end of the dc curve may or may not be significant since it involved measurements at the limit of the sensitivity of the electrometer." Another related note may be found on page 4, *ibid.*; "It can be seen that at the lowest measured temperatures the conductivity becomes almost independent of majority concentration, whereas at higher temperatures the dependence is pronounced." Another note is referenced by the number 21 on page 20, *ibid.*; "In this connection we might add that our experimental activation energies for dc conduction are not in as good agreement with dc theory as the cases treated by Miller and Abrahams. (A. Miller and E. Abrahams, Phys. Rev. 120, 745 (1960).)" We think that, the discussed subject may be related to the inductive contribution for the conductivity, i.e., the parameter for $\mu$ (L) in Eq. (9) here; which may not be (ignored or) taken as infinity. More over, the parameter for L should be $L=(Cw_0^2)^{-1}$ with $w_0 \sim 0.01$; so that, $w_0$ (or a frequency about $w_0$) defines the minimum for the conduction as mentioned in Sec. 2.3 case 2) here. Please note that, L is taken as infinite for the glasses (Sec. 3.2), where we have different mechanism underlying currents at high temperatures. So, a finite parameter for L may be a peculiarity for silicon or for small temperature behavior for all the (disordered) materials. We favor the second choice; since the mentioned glasses and the silicon pronounce similar disordered systems. We suggest that, the number of the magnetic coupling in terms of the flux linkages ($N_{flux}$), i.e., the number of wounding turns is small for $T \sim 0$ $^0K$, since many of the donor electrons are bound yet (where we have polarization and displacement currents mainly). When they ionize for $T \sim 25$ $^0K$ they contribute to the conduction in terms of charge transportation and ($N_{flux}$) increases, and hence L increases since $L \propto (N_{flux})^2$ for any shape (toroid or solenoid, or else).

It is obvious that $\rho$ is very big for $T \sim 0$ $^0K$ and $\rho^{-1}$ may be ignored in $\sigma$; and hence we have only capacitive and inductive contributions for $\sigma$, so that $\sigma_{min}=\sigma(w=w_0) \neq \rho^{-1}$ in the present temperature regime. (We think that, $\mu$ (L) is also big, yet finite for $T \sim 0$ $^0K$) In the following regime; the density of the electrons (free charge carriers) increases with T (Arrheniıs behavior), as a result $\rho$ ($\rho^{-1}$) decreases (increases). In the mean time $\mu$ (L) and $\varepsilon$ (C) couples, since the currents in tiny loops at $T \sim 0$ $^0K$ unify and form many big loops with various radii (which may be considered as the extreme disorder limit; please see Appendix).

Up to us, one direct evidence for $\sigma_{min}=\sigma(w=w_0) \neq \rho^{-1}$ (so, L) is the fluctuation in the experimental $\sigma(w)$ data (plots) for small T (Figures 5-7 here or related plots in the literature; for example, Fig.1 in [16] for glasses). It may be observed that the mentioned fluctuations diminish as T (hence $\sigma$) increases. Secondly, they emerge on every plot for big or (relatively) small w. Thirdly, the amplitudes of the mentioned fluctuations seem as independent of the related $\sigma$ and they correspond to different frequencies which are some multiples of the frequency denoted for the given plot. And finally, there is no explanation on theoretical (or experimental) bases in the literature (up to our knowledge) about the present issue. We suggest the following explanation: i- Any ac for a frequency w is sum of many ac with various frequencies about w. ii- As the quality of the ac supply increases the difference between the mentioned frequencies decreases, yet they may not be totally got rid of for an absolute sinus wave with a unique (and constant) frequency. iii- It is known that the sum (super imposition) of two sinusoidal waves with close frequencies (w and w+w' say) is given as; Sin(w+w'/2)Cos(w'/2), where w'~0. Hence, not an ac for a single w is possible but a wave packet for many frequencies about w; where, the envelope of the packet defines the ac with very small frequencies which are important for finite inductive contributions in terms of $\mu$ (L), as exemplified in Fig. 2 here. Please note that, $\sigma(w) \sim \sigma(w')$ for $ww' \sim w_0^2$ for finite inductances; where, $\sigma(w)$ is symmetric about a vertical line crossing the w-axis at $w=w_0$ (Fig. 3 here). And we have conductivity fluctuations for any w for small T where the inductances are finite.

The mentioned (one paragraph before) fluctuations for small T may clearly be observed in Figure 4 [15] and Fig. 5 here, which is reproduced from Fig. 4 *ibid*. Secondly; the amplitudes of the fluctuations are increasing with w for a given T~0 (T=4 $^0$K). We expect no fluctuation in the electro dynamical parameters ($\rho$, $\mu$, $\varepsilon$, etc.) with T~0 (T=4 $^0$K) and for the experimental frequencies.

In Fig 5 here, we observe that there are three temperature regimes in $\sigma$: i- At about T~0 $^0$K (T< ~ 4 $^0$K) we have a steep increase. ii- For ~ 4 $^0$K <T< ~ $E_A/k_B$ the increase is smooth. iii- For ~ $E_A/k_B$ <T the increase becomes steep. We attribute the case i- to the expected behavior about the absolute zero; case iii- to the expected behavior in terms of Arrhenius temperature dependence; case ii- to thermal expansion, where $\mu$ and $\varepsilon$ change since the physical dimensions of the sample change with T. Please note that, in Figs. 1-4 here, we had considered a material which involved acceptors (Boron) and donors (Phosphorous) with the concentrations $0.8 \times 10^{15}$ cm$^{-3}$ and $2.7 \times 10^{17}$ cm$^{-3}$, respectively; where the effect of varying the physical dimensions in terms of the thermal expansion is negligible. And now, we have small concentrations for the acceptors (undefined) and $1.5 \times 10^{16}$ cm$^{-3}$ for Phosphorous donors. We think that, thermal expansion may be unimportant if the material is doped with acceptors and donors at the same time but if it is doped with acceptors (or donors) at a time. Secondly, we may have important thermal expansion (contraction) for donors (or acceptors).

It is obvious that, we have (indeed) four temperature regimes in Fig. 5 here (experimental, [15]) if the conductivities for big w and T are also considered for $10^{-5}$ $\Omega^{-1}$ cm$^{-1}$<$\sigma$, $10^5$<<w and 20 $^0$K<T as can be observed at the right top corner of Fig. 5 here. And our approach for the present situation will be the following: We utilize Eq. (15) for $\rho$(T) and Eq. (10) for the thermal expansion of the sample with some exponential coefficients and powers for T. It is obvious that, the geometrical factors for the ac current elements (L, R and C) in Eq. (12) are important here for the given $\mu(T_0=1.2\ ^0K)$, $\rho(T_0=1.2\ ^0K)$ and $\varepsilon(T_0=1.2\ ^0K)$. Secondly, we may expect the related exponentials for the g-factors bigger here with respect to these for the material in Figs. 1-4.

Fig. 6 is the theoretical results for $\sigma$ for the sample and the parameters given in Table I (the second row) and Fig. 7 is he same as Fig. 6 for $\sigma$ but as a function of T/10. We predict that the linear dimension (length, radius, width, etc.) of the sample changes about 1.8 % within the present case.

Many other experimental figures in [15] may be described and are mimicked (not shown) in terms of the present model, with some minor changes in the parameters which are given in Table I. We may remark that, as acceptors are doped in n-type Si, the temperature behavior turns to T$^{1+\gamma}$ with 0<1+$\gamma$ from this with 1+$\gamma$<0; and for some concentrations in between we may not have any temperature dependence in $\sigma$(w) (for zero as a power of T). The described transition may be occurring abruptly, since we have $\gamma$=-1 then, and the exponential coefficient becomes very big. Secondly, as the concentration of the majority impurities increases, $\gamma$ increases in the parameters of Eq. (12) here. When the mentioned increase is performed by a factor of ten or so, then log$\sigma$(w) may increase in the same order or more, as we discuss for the mixed glasses in Sec. 3.2. We mimicked (not shown) Fig. 8 in [15] for different majority concentrations in the following way: i) We assumed positive powers for T in the definitions of the parameters for the ac current elements and the geometrical factors in Eq. (12). ii) We assumed big $\gamma$ for high concentrations in the same definitions. We obtained similar results to the experimental ones, where we increased the mentioned parameters in small amounts till we obtain good agreement in between. It is obvious that, we followed the mentioned approach here since we do not have independently defined parameters for the state function. Yet, it is important (up to us) that, the experimental data for the ac could be explained consistently in simple terms of the model, for different systems in different temperature regions, etc. We think that the present approach may be used to discover many thermodynamic properties of disordered solids.

*3.2. Glasses; sodium borates* $xNa_2O.(1-x)B_2O_3$ *and mixed alkalis* $xLiF–(0.8-x)KF–0.2Al(PO_3)_3$

Glasses are known as liquids with high viscosity and the exponential form for thermal expansion (Eq. (10)) may suit well here (also for a highly viscous ionic melt just above the glass transition [8], etc.). Fig. 2 in [16] is the "Arrhenius plot of the dc conductivity of sodium borates of different compositions, $xNa_2O.(1-x)B_2O_3$", where, the horizontal axis is in 1000/T and the vertical one is in $\ln(\sigma_{dc}T)$ and each line is for x between 0.10 and 0.30 in steps of 0.05. We calculate (roughly) the slopes as -14.37, -14.06, -11.25, -9.62 and -7.81 for x as before, respectively.

Figure 8 is the theoretical results (where the present model is utilized) for $\ln(T\sigma(w))$ as a function of 1000/T for a frequency range from $w=10^{-19}$ to $w=10^{20}$, for $xNa_2O.(1-x)B_2O_3$, with x=0.3. The envelope for $\ln(T\sigma(w=10^{-19}))$ gives the so called "activation energy for the dc conductivity" with a slope $\sim -87/11$ in the mentioned glass. The two sided dashed arrow indicates $T_0 = 273\ ^0K = 0^0\ C$. The resulting isotherms are given in Figure 10 (right), for $T\sigma(w)$ with the same range in T and w as Fig.1 in [16]. Figure 9 is the same as Fig. 10 (right) but, for $\sigma(w)$ in a bigger frequency range, where it may be observed that the slope of $\sigma(w)$ is one, asymptotically, in the log-log scale, which is independent of T. Furthermore, the ac conductivity is (almost) independent of w in the valley. This property, which is common for many solids, may be explained as follows: If only $\rho^{-1}$ increases (decreases) in Eq. (9), then $\sigma(w)$ increases (decreases) only in the valley and in the gradual transition region for intermediate w. Same is true with ε and in the right (C-branch) and in the gradual transition region. In terms a plot similar to Fig. 8, where $T\sigma(w)$ is plotted with respect to w in log-log scale (not shown), we predict that the slope for big w increases as T or w increases. Therefore, we can conclude that, the increase in T is reflected to the increase in $\sigma(w)$ via the resistive (capacitive) term for the valley (C-branch), and since the term in the exponent ($\alpha_C \sim \alpha_L$) is smaller than $\alpha_R$ (Table II), the increase in the C-branch is minor. It is obvious that we do not see the effect of inductance here, since its parameter comes out semi empirically as very big (Table II) and $w_0$ is very small ($\sim 10^{-25}$).

In high temperature regimes, 1/T terms with arbitrary powers in Eq. (10) become less important; and hence the temperature dependences of the related parameters simplify. As a result, many scaling properties may emerge in the conductivity, which are local i.e., about some temperature (T' say) and approximately. For example, the expression $(\sigma(w) – 1)\rho$ in Eq. (11), which is dimensionless, may yield a scaling, since the inductive term may be omitted for big T. Then we have $(\sigma(w) – 1)\rho= \rho Cw$, where the exponent of the multiplication $\rho C$ may be expanded as (approximated with) an expression linear in T about some temperature (T', say) within the experimental range. As a result, it may be showed that the conductivity displays the following scaling property: $(\sigma(w) – 1)\rho \cong (\rho'C')w + (d(\rho C)/dT) (T-T')w= Uw + uTw$, where $U=(\rho'C') - T'u$ and $u=(d(\rho C)/dT)$ which is evaluated at T=T'. More over, U and u may be similar (or the same, approximately) for a group of (different) materials, then we have the mentioned scaling (or the so called "universality") for the mentioned solids, etc. Next, suppose that $d(\log_e(T\sigma)/d(1/T))=s$, then it can be showed that $\log_e(\sigma/\sigma')= – \log_e(T/T') + (s/T) – (s/T')$, where $\sigma'=\sigma(T=T')$. Thus $\sigma(T) \rightarrow (T/T')\sigma(T)$ and we have another scaling property. The important point here is that, the results for several scaling (empirical) may be used to define the model parameters empirically.

The isotherms in Fig. 10 (with x=0.3, right) are similar to these in Fig. 2 in [16] with the same temperatures. (Please see also Fig. 11 here, for the alkali glasses.) Please note that, $R(T=T_0)$ and $C(T=T_0)$ are approximately constant for $0.1<x<0.3$. We claim that, increasing

Na$_2$O concentration in the glasses decreases the so called activation energy, i.e., the coefficient for the thermal expansion which helps for the expansion, and vice versa.

Figure 11 is for the mixed alkali glasses; xLiF–(0.80-x)KF–0.20Al(PO$_3$)$_3$ with x=0.5 and is similar to Fig. 1 in [3]. The related parameters (Table III) are empirically defined in the same manner as before. Collapsing the data for several scaling in terms of x, R, T, etc., is trivial, as described two paragraphs before. Secondly; we think that, the difference is big in the exponential coefficients of the given ρ(T) (σ$_{dc}$(T) in [3]) for the starting materials (0.80KF–0.20Al(PO$_3$)$_3$ and 0.80LiF–0.20Al(PO$_3$)$_3$); so that, when the materials are mixed, then the mentioned coefficients may also mix with the same x (linear interpolation), which gives a quadratic behavior (almost) in ρ(T); and hence ρ$^{-1}$(T) becomes the extreme (minimum, here) at about x=0.5. As a result the ac conductivity can not be collapsed on a single plot, but two for various x; Figs. 3 and 4 in [3] (the so called mixed alkali effect (MAE), *ibid*.).

The mentioned turn over in ρ$^{-1}$(T) (or σ$_{dc}$(T), as used in the literature) for mixed alkali glasses could be found in sodium borates as well. But, a small regime in x (0.1<x<0.3) is considered in [16] and the borates were always dominant there. As a result, the coefficients in the related exponents come out as linear in x; and, we think that the similar turn over in σ$_{dc}$(T) for sodium borate glasses could be met for x~0.5. As a result, the experimental data for σ(w) for various x could not be collapsed into a single plot, but two.

Please note that, the experimental data (for the mentioned glasses) may be mimicked well by the present model and many "mysterious" properties related to scaling and "universality" may be resolved into simple terms. Further more, several aspects of the phenomenon may be predicted. For example, we predict that the conduction mechanisms are similar for the mentioned glasses here and in doped Si for big T (Sect. 3.1), etc.

## 4     Discussion and conclusion:

The present model is applied to different ac solids for different frequency and temperature regimes (1 Hz-10$^6$ Hz or big and 1 $^0$K-500 $^0$K, respectively) and the out coming results of the model are found in good agreement with the experiments (Sec. 3). It is obvious that the present model treats the frequency dependence of the electrical conductivity in terms of the classical theory of electromagnetism (Maxwell equations) and the temperature behavior in terms of the thermo dynamical state function (Secs. 2.1 and 2), where the underlying mechanisms (for both the electrical and thermal conduction) are well known to be the same. Secondly, the model predicts several scaling properties for a given (or a group of) material(s), where the mentioned properties come out as a result of the similarities in the exponential behaviors of the related parameters in terms of various powers for the temperature; the coefficients and power terms in T are decisive for different temperature regimes in the conductivity.

The displacement and polarization currents are underlying the conductivity for the cases where resistive (Ohmic) conductivity is not present or negligible: For high T or small T (~0 $^0$K) and for big w (w$_0$<w) in all. At intermediate temperature regions (T~E$_A$/k$_B$) resistive (Ohmic) conductivity becomes important and hence the minimum conductivity σ$_{min}$(w~w$_0$) involves a term for resistive contribution, besides the capacitive and inductive ones.

For the case with low T or small w, the g factor in Eq. 9 is crucial, which pronounces the size effect since the mentioned factor is predicted as proportional to the length of the current path (which is greater than or equal to the length or thickness of the sample) in the inductance per cross sectional area (for the same current). Hence, the conductivity may be bigger for thin and long samples with respect to wide (flat) and short samples for the same material. We think that the g factor and hence the inductive contributions become negligible for high T and high (extreme) disorder, where the inductive currents (and the inductances) unify and we have the whole solid as a single inductance, i.e., solenoid. Please note that, the length of a current path in a solenoid may be estimated as $(\delta^2 + (2\pi rN)^2)^{1/2}$, where N is the total number of turns for wounding and r is the radius (on the average) of each turn, so that the cross sectional area (a) of the solenoid is $a=\pi r^2$. Hence, the g factor (Eq. (9)) may be estimated as $(\delta^2 + (2\pi rN)^2)^{1/2}/\pi r^2$, where $r^2 \sim N'$ for some N' with NN'=M if the total number of the disorder centers in the solid is M. We suggest two evidences (besides Maxwell equations and the equipartition of energy theorem as discussed in Sec. 2 case 5) which indicate the importance of the inductance for low T and small w: i- The minimum conductivity does not follow $\rho^{-1}$ here. ii- We have several fluctuations in $\sigma(w)$ for a given experimental w, where the inductive contributions emerge for small frequencies about $w_0^2/w$ in terms of the envelopes of the experienced wave packet in terms of the sinus waves with close frequencies about w. Yet, there is not much information about the subject in the experimental literature and we do not consider the inductances to further extend. Secondly, they are important for low T and small w regimes which are not covered in the majority of the available experimental reports.

It may be remarked that, the complete treatment of the inductive contributions for the conductivity in theoretical terms could be meaningful if the experimental data for the ac solids in wider T and w regions were available.

**Appendix**

Random networks are known to be useful to model disordered solids, where the resulting conductivities are obtained in terms of simulations.

Indeed, we had applied a random LRC network initially; where, the ac circuit elements (L, R and C) are uniformly random distributed with various concentrations. The voltage differences between the ends of arbitrary current elements (sites) are considered for arbitrary current paths; by summation (integration) of the individual contribution (to the ac conduction) of each element we obtain the space (time) average over many sites (periods) for the current versus voltage plot as a function of frequency (w) for the ac conductivity $\sigma(w)$, i.e., the so-called "master" curve. (Up to our knowledge, there is no experimental assertion about the dimension of the current paths in solids. Yet, there is a note in [18] (page 3) stating that "Computer simulations have shown that at extreme disorder the dc current follows almost one-dimensional paths (Brown and Esser, 1995)" [20]. In any case, the present approach could be considered as an assumption. Yet the results for the ac conductivity come out for a single element, where the parameters for the related ac circuit elements are defined uniquely for the given solid.)

We define a uniformly distributed random number ($\xi$) between zero and 1 for each site ($r_n$) at any tour (t) of the simulation, and take $r_n$ as an inductive site if $1 - c_L \leq \xi < 1$, or a capacitive term if $0 \leq \xi < c_C$, or resistive one if $c_C \leq \xi < 1 - c_L$;

$$X_n = X(r_n) = \begin{cases} X_L & \text{if } 1-c_L \leq \xi < 1 \\ X_R & \text{if } c_C \leq \xi < 1-c_L \\ X_C & \text{if } 0 \leq \xi < c_C \end{cases}, \quad (A1)$$

where, $c_R = (1-c_L) - c_C$ obviously.

Secondly, we consider the LRC network not as a whole but in terms of current lines ($I_m$); and then, we consider each current line (branch, path) in terms of individual current elements (segments): A current element at any site ($r_n$) at any time (t), $I_{mn}(r_n, t)$ is part of the current path ($I_m(t)$). It is known that the current must have the same amplitude and the same phase (in a series LRC circuit) for any $n \leq N_m$, $m \leq M$ at any t, where $N_m$ and M are the number of the individual current segments (along the current path, m) and the number of the current paths in the solid, respectively.

Finally, for the ac with a given frequency w; we take $X_L$ or $X_R$ or $X_C$ (=wL or R or 1/wC, respectively) and $X_n$ (A1) becomes frequency dependent ($X_n=X_n(w)$). Then we calculate the voltage differences (for I versus V plots) $V_{mn}(r_n,w)$ between the ends of each segment (site); and, we take the average of $V_{mn}(r_n,n)$ over n and m (and t) for the ac conductivity $\sigma(w)$;

$$V_{mn}(r_n,w) = I_{mn}(r_n) X_n(w) \quad ,$$

$$V_m(w) = (1/N_m)\sum_n^{N(m)} V_{mn}(r_n,w)$$

and
$$V(w) = (1/M)\sum_{N(m)}^{M} V_m(w) \quad . \tag{A2}$$

One may utilize similar random numbers ($\xi$, (A1)) for the (definition) distribution of the parameters for L, R and C over the network entries ($r_n$) for any t;

$$L(n,t) = (L_2-L_1) \xi + L_1 \quad , \tag{A3}$$

where, different $\xi$ is tried for each ($r_n$) at any t. In (A3) $L_2$ is the maximum value and $L_1$ is the minimum value for L; and, it is obvious that, one may select $L_1=0$ or $L_2=L_1$, etc., and all similarly for the definition of R and C. (It may be worth to emphasize that, if no randomness is assumed in the given parameters, then $L_2=L_1$ ($R_2=R_1$, etc.) may be selected in A3.)

It is obvious that, the LRC network may not match with the realistic network of the atoms in a solid. Secondly, we do not know whether our paths run over the ac circuit elements which are connected in series or in parallel (or in a more complicated way). The elements are all represented by the parameters which are uniformly random distributed over the segments, for each path. And, thus; all the alternating current circuit elements may be considered (as if) connected in series (with random parameters; (A3)), along each single path.

The averages taken over time or over (m) are equal (A1). Similarly, taking the average over one very long path ($N_1 \rightarrow \infty$) amounts to taking average over many shorter paths (with $N_1=N_mM$).

We discuss several reasons in the main text, for taking the inductances into account for the conductivity and we utilize LRC network to model a solid, instead of RC.

The subject of the constant amplitudes and phases over the current elements may be considered as an assumption or a selection for the model: If the out coming current amplitudes and the phases for the individual segments are not the same (due to any reason, say temporarily), then their average for big $N_m$ and M and over a finite time period is zero; which means that, this path is closed for the conduction, and we disregard it. And we select only the current carrying paths.

The definition of each parameter (for L, R and C; (A3)) for individual sites involves geometrical terms such as length, radius, cross sectional area, etc. Therefore, some atomic distances (as the bond length for polymers and amorphous materials (in average) or the lattice constant for crystalline solids; all few Angström ($10^{-10}$m)) may be used in attempts to calculate (estimate) the related definitions in microscopic approach. Or, some other approaches, such as the macroscopic view or the empirical results (if available) may be utilized on the same purpose.

Finally, we use exponential expressions for the temperature (T) behavior of the parameters L, C and R, where several terms in various powers of T were considered (as in the related thermo dynamic state function) and the coefficients were obtained semi empirically or in terms of try and error. The out coming results for the simulations were in good agreement with the experimental data for Si and several glasses as in Sects. 3.1 and 2.

We started with a network and came up with a simple expression for a single LRC circuit for the solid; where, the individual LRC circuit elements may be connected (neither in series nor in parallel, but) in a different way, so that the average of the reciprocal of the reactance for each term defines the conductivity σ(w). It is obvious that, Eq. (9) in the main text can be obtained from (A2) here, within the following approach: The modeling current paths run along the direction of the applied current (i.e., the electric field) with some probability, say $p_1$ or they run with a different probability over the aside sites which are in perpendicular direction for the applied current, say $p_2$. It is obvious that $p_1/p_2 \propto \delta/\Delta$, where δ is the width (the length for the applied ac voltage and the current) and Δ is the cross sectional area of the solid (where the volume of the sample is Δδ, obviously). Please note that Δ/δ (or δ/Δ) defines the geometrical factor for the ac circuit elements, as discussed in Sec. 2.3. Please note that, Eq. (9) is obtained in terms of the classical theory for the electromagnetism (Maxwell equations) through Eqs. (1)-(9) and (11)-(12) in the main text.

**Acknowledgments**

The author is thankful to Prof. Dr. Bayram Katırcıoğlu (METU) for mentioning the subject some months ago.

**TABLES**

**Table I**: The model parameters (Eqs. (9), (10) and (12)) for the n-type Silicon samples in [15](*)

| Sample | $\mu(T_0)$ | $\alpha_L$ | $\gamma_L$ | $\rho(T_0)$ | $\alpha_R$ | $\gamma_R$ | $\varepsilon(T_0)$ | $\alpha_C$ | $\gamma_C$ | $w_0$ |
|---|---|---|---|---|---|---|---|---|---|---|
| 1 (13) | $10^{15}$ | -3.4 | -2.0 | $10^{28}$ | -34 | -2 | $23.5 \times 10^{-14}$ | -3.4 | -2.0 | 0.2 |
| 2 (8) | $10^{18}$ | -2 | 0.001 | $10^{30}$ | -30 | -2 | $23.5 \times 10^{-16}$ | -2 | -0.001 | 0.02 |

(*) Sample numbers in [15] are repeated in parentheses here. Sample 1 (13) is for Figs. 1-4 and 2 (8) is for Figs. 6 and 7. $T_0 = 1.2\ ^0K$ (Eq. (10)) and the g-factors (Eq. (12)) are absorbed in $\mu(T_0)$ in all.

**Table II**: The model parameters (Eqs. (9), (10) and (12)) for the sodium borate glasses $xNa_2O \cdot (1-x)B_2O_3$ with $0.1 < x < 0.3$ [16] (*).

| Parameter | x=0.10 | x=0.15 | X=0.20 | x=0.25 | x=0.30 |
|---|---|---|---|---|---|
| $\alpha_L$ | 0.055 | 0.054 | 0.043 | 0.037 | 0.03 |
| $\rho(T_0)$ | 0.0184 | 0.018 | 0.0144 | 0.0123 | 0.01 |
| $\alpha_R$ | -1.437 | -1.406 | -1.125 | -0.962 | -0.781 |
| $\varepsilon(T_0)$ | $1.84 \times 10^{-10}$ | $1.8 \times 10^{-10}$ | $1.44 \times 10^{-10}$ | $1.23 \times 10^{-10}$ | $10^{-10}$ |
| $\alpha_C$ | 0.055 | 0.054 | 0.043 | 0.037 | 0.03 |

(*) $L(T_0) = 10^{40}$ and $\gamma_L = \gamma_R = \gamma_C = \sim 0$, for all; $T \rightarrow T/1000$ is performed for small exponents for the software and $T_0 = 273^0$ C (Eq. (10)).

**Table III**: The model parameters (Eqs. (9), (10) and (12)) for the mixed alkali glasses $xLiF–(0.80-x)KF–0.20Al(PO_3)_3$ with $x=0.5$ [3] (*)

| parameter | x=0.5 |
|---|---|
| $\alpha_L$ | 5.6 |
| $\rho(T_0)$ | $1.25 \times 10^{-3}$ |
| $\alpha_R$ | -23.412 |
| $\varepsilon(T_0)$ | $5 \times 10^{-9}$ |
| $\alpha_C$ | 5.6 |

(*) $L(T_0) = 10^{40}$ and $\gamma_R = 1.0$, $\gamma_L = \gamma_C = 0.2$, for all; $T \rightarrow T/1000$ is performed for small exponents for the software and $T_0 = 473^0$ C (Eq. (4)). Parameters for other x could not be obtained due to lack of the experimental data in [3].

**FIGURES**

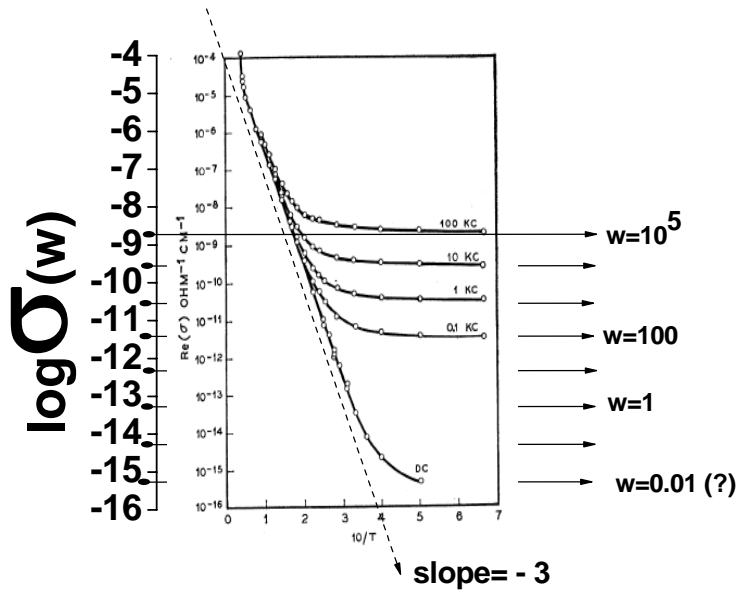

**Figure 1**    Reproduction of Fig. 5 in [15] for the inverse temperature dependence of the ac conductivity and the dc conductivity in n-Si with Boron (acceptor) impurities (concentration= $0.8 \times 10^{-15}$ cm$^{-3}$) and Phosphorus (donor) impurities (concentration= $2.7 \times 10^{17}$ cm$^{-3}$) as given (sample number=13, crystal number 215) in Table I, *ibid*.

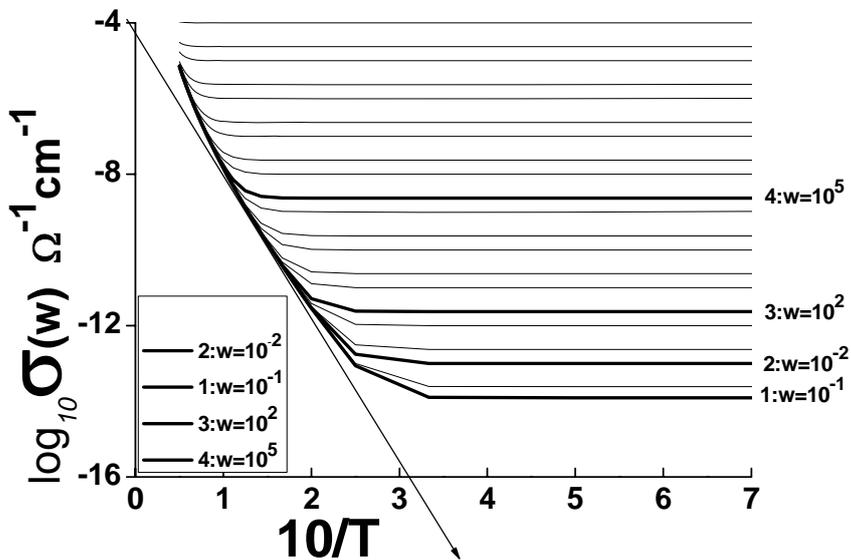

**Figure 2**    Same as Fig. 1, with the model parameters given in Table I, where Eq. (15) is utilized for the resistivity (please see Sect. 3.1).

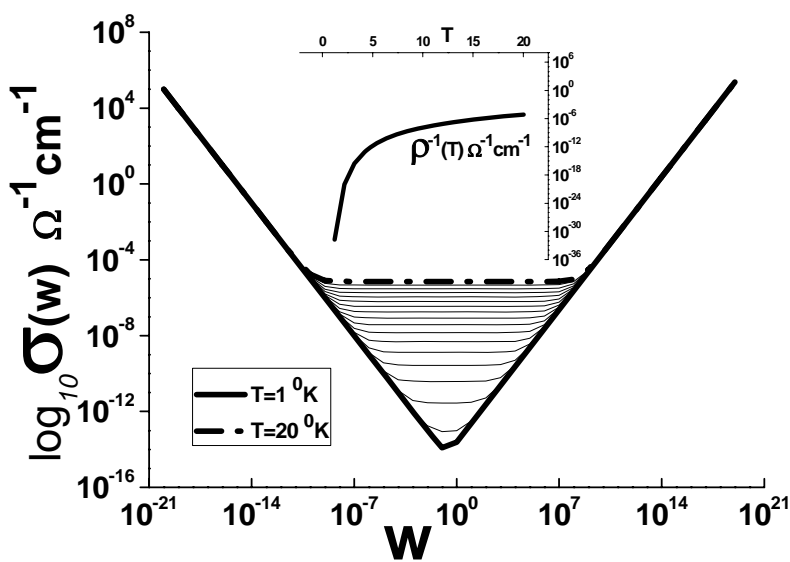

**Figure 3**   Isotherms for the conductivity of the same material as in Fig. 1, where the minimum of the conductivity is at $w_0=0.2$ Hz. Please note that the conductivity is symmetric with respect to the vertical line passing through $w_0$. Inset is for $\rho^{-1}(T)$; please see Fig. 4.

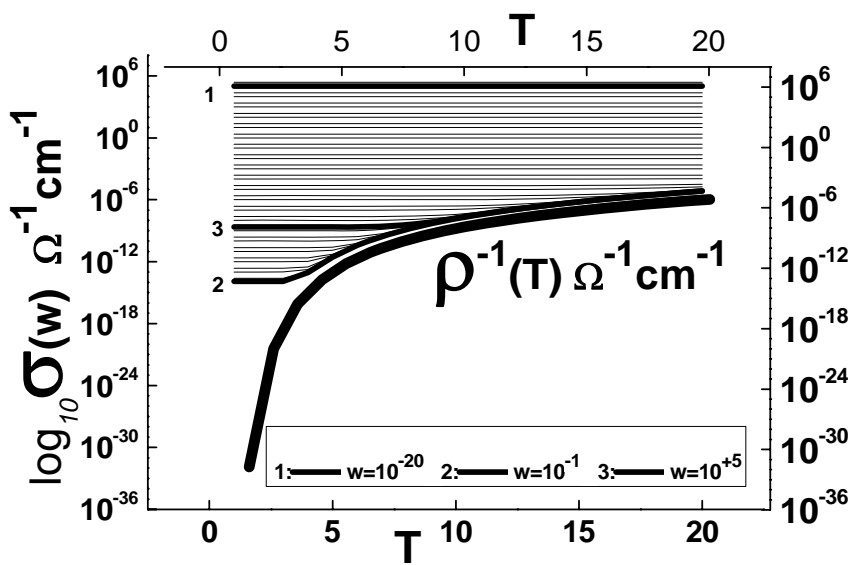

**Figure 4**   The conductivity as a function of T and w, versus $\rho^{-1}(T)$; where, the minimum conductivity at any T is defined by $\sigma(w=w_0=0.2)$. Please note that the lines denoted as 1, 2 and 3 for the conductivity are for $\sigma(w_1=10^{-20})$, $\sigma(w_2=10^{-1})$ and $\sigma(w_3=10^{+5})$, respectively (all thick); and the thin lines are for other frequencies. Secondly; the axes for $\rho$ are (top and right) little shifted with respect to these for $\sigma$ (bottom and left) intentionally.

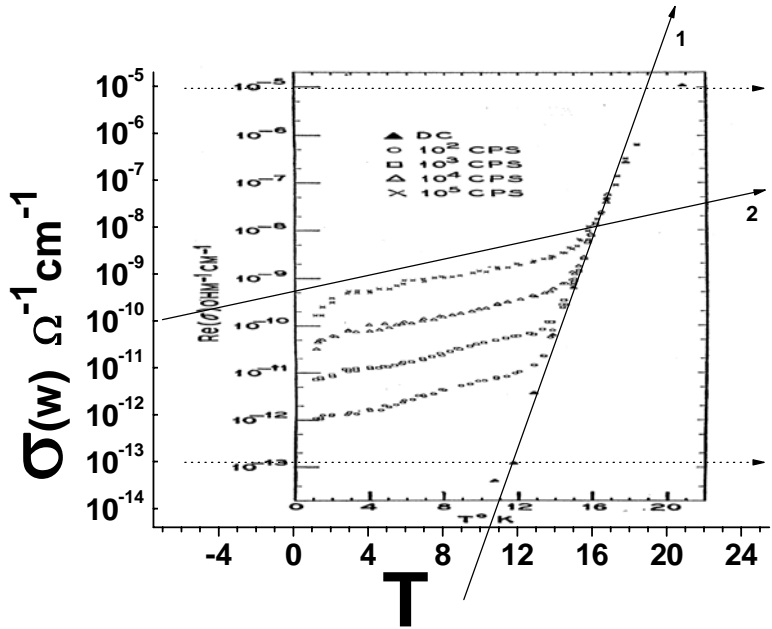

**Figure 5** The experimental conductivity for doped Si (for the sample 8; described in Table I [15]). The present figure is reproduced from Fig. 4 *ibid.*, where the arrows 1 and 2 (tilted) designate different temperature behaviors and the dotted (horizontal) arrows are for matching the scales in the axes. Please note the fluctuations in the conductivity for various w and T.

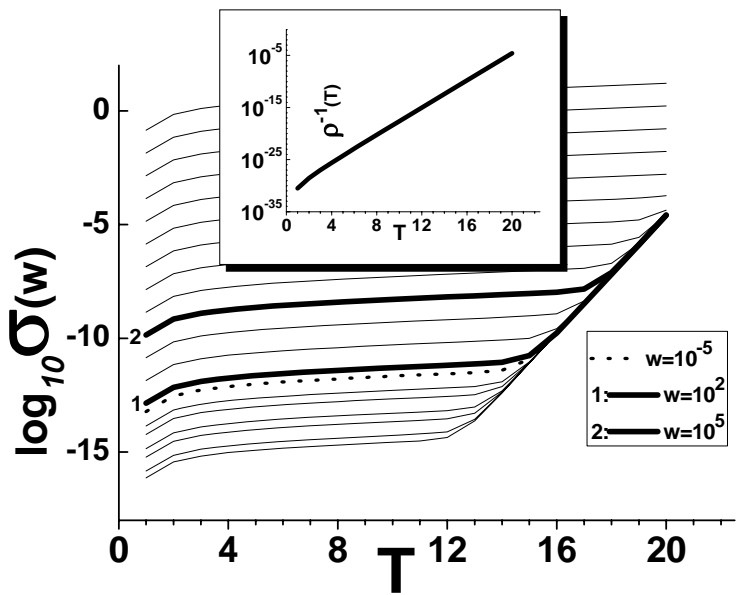

**Figure 6** The theoretical conductivity for the doped Si as in Fig. 5. Please note that $\sigma(w)$ and $\sigma(w')$ are nearly equal for $ww'=w_0^2$, where $w_0^2=0.02$ for the parameters on the second row of Table I. Secondly, the lowest plot is for w=0.01, which is not the dc conductivity; $\rho^{-1}(T)$ is shown in the inset which approximates $\sigma_{min}(w)$ for all the frequencies at 12 $^0$K<T; for smaller temperatures we have polarization and displacement currents.

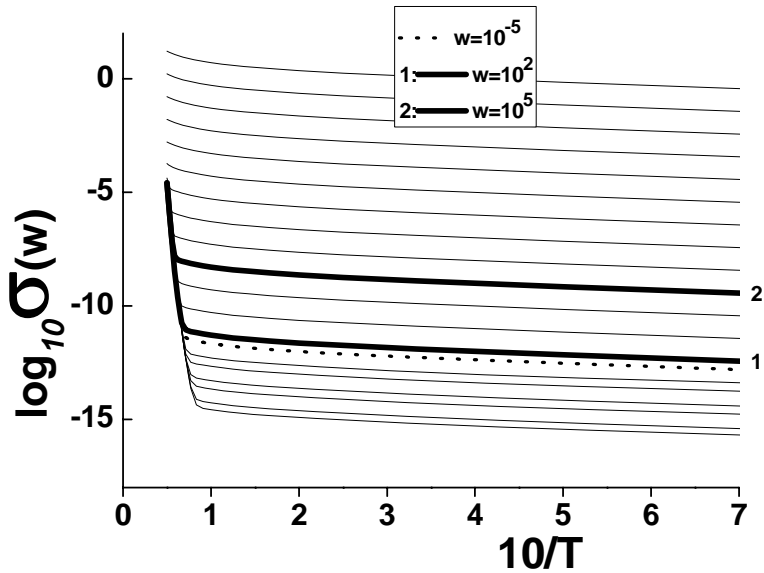

**Figure 7**    The theoretical conductivity the same as in Fig. 6 but as function of 10/T.

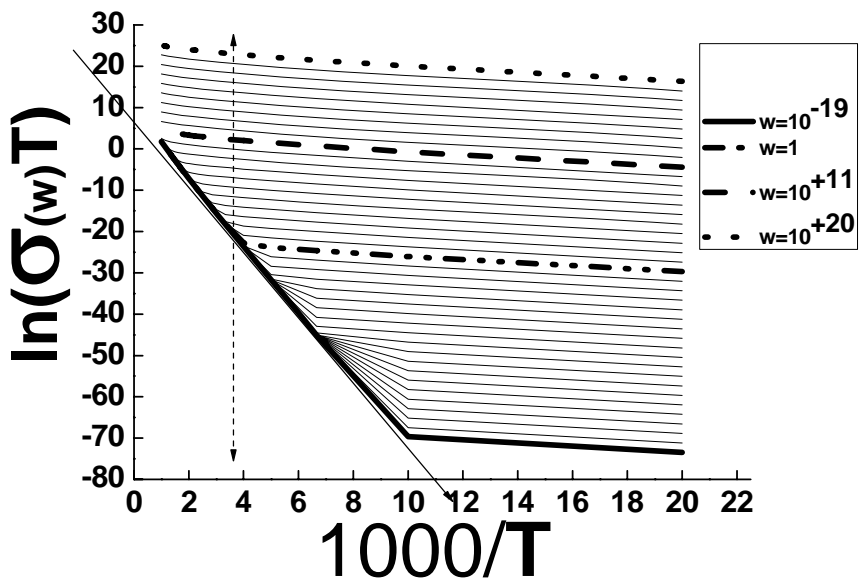

**Figure 8**    Results of the present model for $\ln(T\sigma(w))$ as a function of 1000/T for a frequency range from $w=10^{-19}$ to $w=10^{20}$, for $xNa_2O\cdot(1-x)B_2O_3$, with $x=0.3$. The envelope or $\ln(T\sigma(w=10^{-19}))$ gives the so called "activation energy for the dc conductivity" in the mentioned glass, which is designated by a tilted arrow (which has the same slope $\sim -87/11$). The two sided dashed arrow indicates $T_0=273\ ^0K=0^0\ C$ for the model parameters in Eq. (5). Please note that $w=10^{11}$ defines approximately the upper limit for conductivity (the plot in dashes and dots).

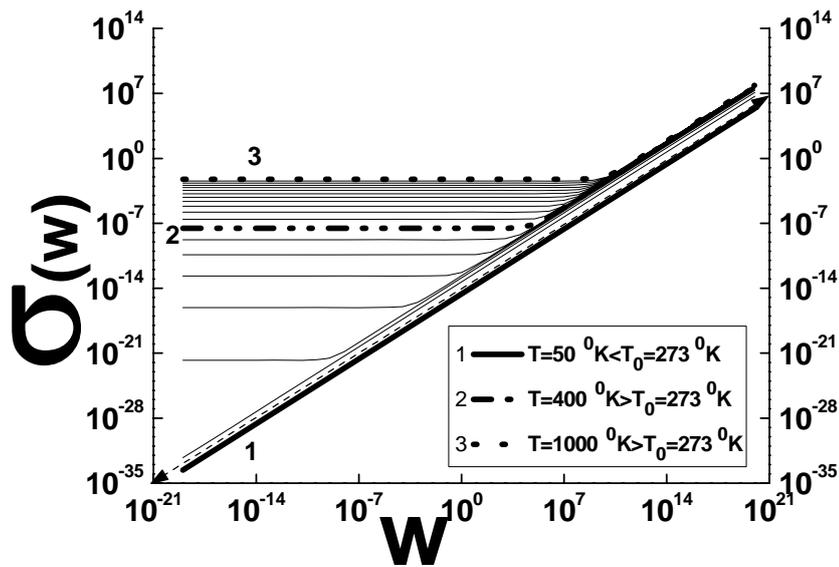

**Figure 9** The isotherms for the conductivity in $xNa_2O \cdot (1-x)B_2O_3$, with $x=0.3$ for T ranging from 50 K$^0$ to 1000 K$^0$ in equal steps, where the two sided dashed arrow indicates the slope of unity in the present log-log scale.

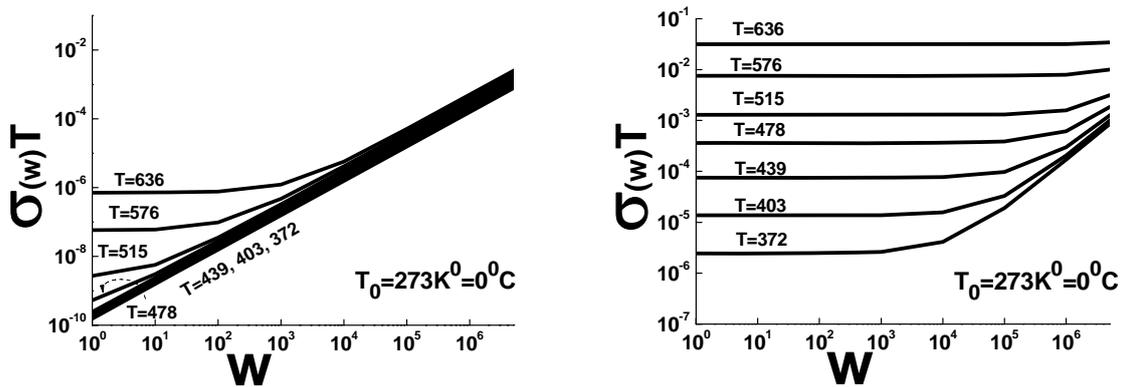

**Figure 10** The isotherms for the conductivity in $xNa_2O \cdot (1-x)B_2O_3$, with $x=0.1$ (left) and with $x=0.3$ (right) for T=372, 403, 439, 478, 515, 576, 636 all in $^0$K, where the parameters of Table II are used. Please note that, the figure at right (for x=0.3) is similar to Fig. 2 in [16] with the same temperatures.

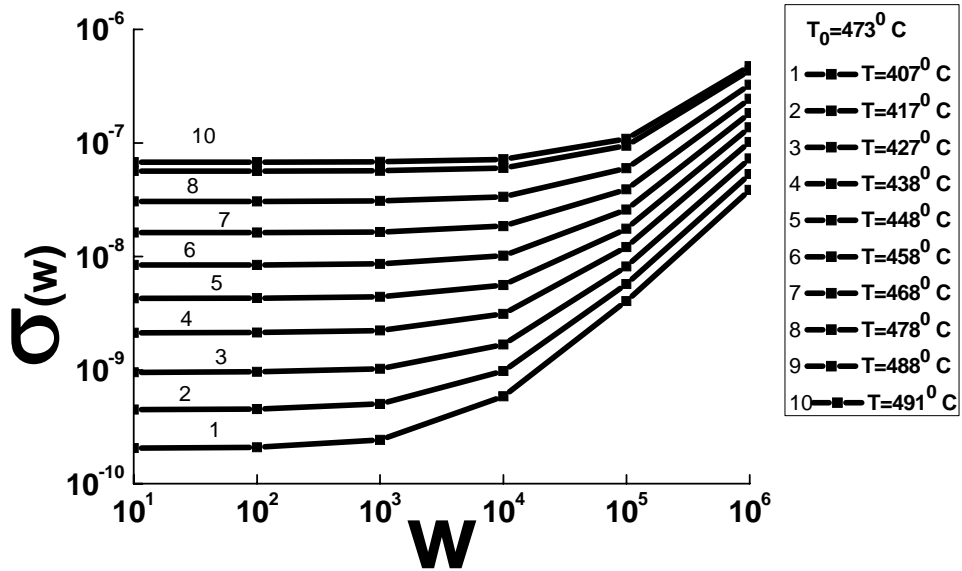

**Figure 11** The isotherms for the ac conductivity in the mixed alkali glasses; xLiF–(0.80-x)KF–0.20Al(PO$_3$)$_3$ with x=0.5 for various T (listed in the legend) where the parameters of Table III are used. Please note that the present figure is similar to Figure 1 in [13] for the same temperatures and the frequencies.